# Topology-preserving digitization of n-dimensional objects by constructing cubical models

Alexander V. Evako
Npk Novotek, Laboratory of Digital Technologies
Volokolamskoe Sh. 1, kv. 157, 125080, Moscow, Russia
Tel/Fax: +7 499 158 2939, e-mail: evakoa@mail.ru

**Abstract**
This paper proposes a new cubical space model for the representation of continuous objects and surfaces in the n-dimensional Euclidean space by discrete sets of points. The cubical space model concerns the process of converting a continuous object in its digital counterpart, which is a graph, enabling us to apply notions and operations used in digital imaging to cubical spaces. We formulate a definition of a simple n-cube and prove that deleting or attaching a simple n-cube does not change the homotopy type of a cubical space. Relying on these results, we design a procedure, which preserves basic topological properties of an n-dimensional object, for constructing compressed cubical and digital models.

## 1 Introduction

Topological properties of two-and three-dimensional image arrays play an important role in image processing operations. A consistent theory for studying the topology of digital images in n dimensions can be used in a range of applications, including medical imaging, computer graphics and pattern analysis. Integrating topological features into discretization schemes in order to generate topologically correct digital models of anatomical structures is critical for many clinical and research applications, where particular regions of the object require a fine resolution while a relatively coarse resolution can be used over the rest of the object of interest (see [16, 20]).
In recent years, there has been a considerable amount of works devoted to building two-, three- and n-dimensional discretization schemes and digital images.
The marching cubes algorithm for extracting a triangulated surface from a regular grid was developed by Lorensen and Cline. Paper [13] contains a survey of this algorithm and its extensions. A variety of representations of n-dimensional meshes have been studied in the literature (see e.g., [19]).
In paper [4], discretization schemes are defined and studied that allow us to build digital models of 2-dimensional continuous objects with the same topological properties as their continuous counterparts. Usually, digital objects are represented by graphs whose edges define nearness and connectivity (see, e.g. [3, 8, 18]). Transformations of digital objects preserve topological properties while changing the geometry of objects. One of the ways to do this is to use simple points which can be deleted from this object without altering topology. The notion of a simple point was introduced by Rosenfeld [17]. Since then due to its importance, characterizations of simple points in two, three, and four dimensions and algorithms for their detection have been studied in the framework of digital topology by many researchers (see, e.g. [2, 5, 9, 14-15, 17]). Simplicial and cubical complexes have been used as tools for building digital images (see e.g. [1-2, 14-15]). In paper [14], elements of a digital picture are associated with unitary cubes instead of points and techniques from cubical homology are applied to the cubical complex. Deleting a simple point means the removal of a unitary cube without changing the topology of the cubical complex. This approach allows characterizing a simple point in terms of the homology groups of the neighborhood of a point. In papers [7, 10-12], properties of contractible transformations of graphs based on deleting and gluing simple points and edges and preserving global topology of graphs as digital spaces were studied. In particular, it was shown that contractible transformations retain the Euler characteristic and homology groups of a graph.

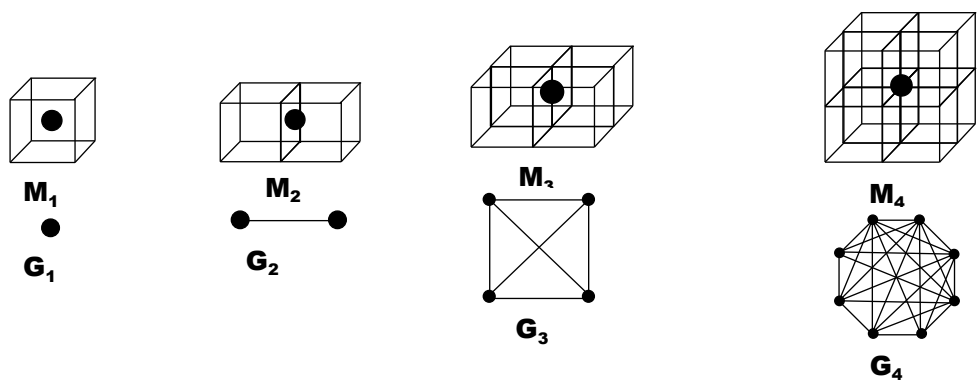

Figure 1. Cubical models for a ball and their intersection graphs. $G_2$, $G_3$ and $G_4$ are homotopy equivalent to $G_1$.

The present paper is dedicated to the study of connection between nD objects, cubical spaces modeling the objects and digital models of the objects. The proposed procedure preserves topological features and allows obtaining digital models of the objects at any required resolution . A cubical space is a key ingredient for this approach. For a given nD object, we first construct a cubical model of the object and then build the digital model of the object as the intersection graph of the cubical model.

The outcome of the paper is based on computer experiments [see [11]) with cubical spaces described in section 2. It follows from computer experiments that a given object S and its cubical model M(S) are homotopy equivalent spaces in the framework of algebraic topology. The benefit of using such spaces is that they establish a deep and direct link between topological properties of continuous n-dimensional objects in $E^n$ and their digital models.

Section 3 presents main results relating to contractible graphs, simple points of graphs and contractible transformations of graphs. Section 4 introduces the notion of a cubical space and a simple n-cube of a cubical space. We show that removing or attaching a simple n- cube does not change the homotopy type of a cubical space. It is shown that if a cubical space is contractible then it can be transformed to single n-cube by sequential deleting simple n-cubes. We prove that the digital model of a contractible cubical space is a contractible graph, and if cubical spaces are homotopy equivalent then their intersection graphs are homotopy equivalent. We describe a method for constructing cubical and digital models preserving basic topological properties of continuous objects.

## 2  Computer experiments

The following surprising fact was observed in computer experiments described in [11]. Suppose that S is an object in Euclidean space $E^n$. For example, it can be an m-dimensional manifold, m≤n. Divide $E^n$ into the set of n-cubes with the side length L and vertex coordinates in the set X={ $Lx_1,…Lx_n$: $x_i \in Z$}. Call the cubical model of S the family $M_1$ of n-cubes intersecting S, and the digital model of S the intersection graph

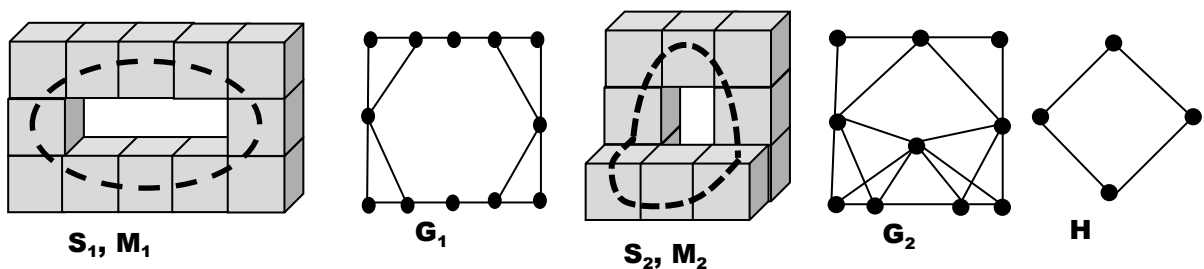

$S_1, M_1$     $G_1$     $S_2, M_2$     $G_2$     H

**Figure 2.** $S_1$ is a circle. $M_1$ is the cubical model for $S_1$. $G_1$ is the intersection graph of $M_1$. $S_2$ is a circle. $M_2$ is the cubical model for $S_2$. $G_2$ is the intersection graph of $M_2$. $G_1$ and $G_2$ are homotopy equivalent to H.

$G_1$ of $M_1$. Then reduce the size of the n-cube edge from L to L/2 and repeat this operation using the same structure. We obtain the cubical space $M_2$ and the digital model $G_2$. Repeating this operation, we obtain the sequence of cubical spaces $M_1, M_2, ...,M_k,…$ for the S. It is revealed that the number s exists such that for all k, p, k>s, p>s, cubical spaces $M_k$ and $M_p$ are homotopy equivalent, and graphs $G_k$ and $G_p$ are digitally homotopy equivalent, i.e., can be transformed from one to the other with contractible transformations.

Moreover, if we have two objects $S_1$ and $S_2$, which are homotopy equivalent, then their cubical models M and N are homotopy equivalent and their intersection graphs G(M) and G(N) can be transformed from one to the other by the same kinds of transformations if the division is small enough.

It is possible to assume that the digital model contains topological and perhaps geometrical characteristics of an object S. Otherwise, the digital model G of S is a digital counterpart of S [11-12].

To illustrate these experiments, consider examples depicted in fig. 1, 2 and 3. In fig. 1, S is a ball, $M_1$, $M_2$, $M_3$ and $M_4$ are sets of cubes intersecting S, and $G_1$, $G_2$, $G_3$ and $G_4$ are their intersection graphs. $M_1$, $M_2$, $M_3$ and $M_4$ are homotopy equivalent spaces in the framework of algebraic topology, $G_2$, $G_3$ and $G_4$ can be converted to $G_1$ by contractible transformations. For a circle $S_1$ shown in fig. 2, $M_1$ is a set of cubes intersecting $S_1$, and $G_1$ is the intersection graph of $M_1$. For a circle $S_2$, $M_2$ is a set of cubes intersecting $S_2$, and $G_2$ is the intersection graph of $M_2$. Obviously, cubical models $M_1$ and $M_2$ are homotopy equivalent, graphs $G_1$ and $G_2$ can be transformed with contractible transformations to H, which is a minimal digital 1-dimensional sphere [5, 12]. A topological sphere $S_3$ in fig.3 is the surface of some cube, $M_3$ is a set of cubes intersecting $S_3$, and $G_3$ is the intersection graph of $M_3$. Clearly, $M_3$ can be continuously deformed $S_3$, and $G_3$ can be converted to $H_3$ by contractible transformations. $H_3$ is a minimal digital 2-dimensional sphere.

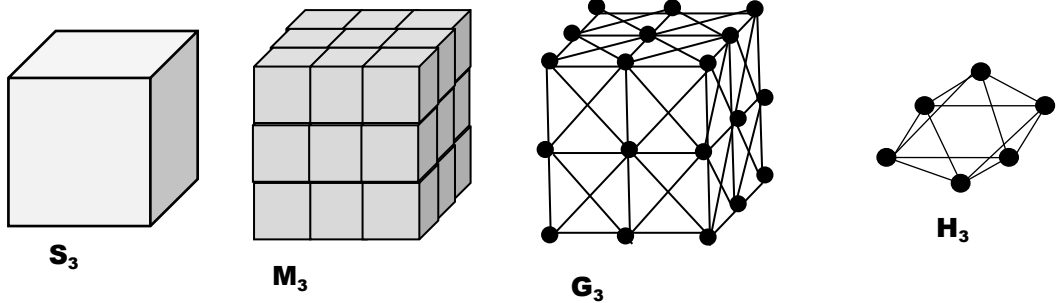

**Figure 3.** $S_3$ is a topological sphere. $M_3$ is the cubical model of $S_3$. $G_3$ is the intersection graph of $M_3$. $H_3$ is homotopy equivalent to $G_3$.

As it follows from above, a cubical model of an n-dimensional object S is a basic structure which allows to obtain the digital model of the object:

- In all cases investigated in computer experiments, a given object S and its cubical model M(S) are homotopy equivalent spaces in the framework of algebraic topology provided that the size of n-cubes is small enough.
- The intersection graph G(M(S)) of M(S) is a natural digital model of S having the same topological characteristics as S.

**Simple points and contractible transformations of graphs**

In order to make this paper self-contained, we include in this section some results related to contractible transformations of graphs. Proofs of basic properties of contractible graphs and contractible transformations can be found in papers [5-8, 10-12]. By a graph we mean a simple undirected graph G=(V,W), where V={$v_1,v_2,...v_n,...$} is a finite or countable set of points, and W = {$(v_p v_q),....$}⊆V×V is a set of edges. Since in this paper we use only subgraphs induced by a set of points, we use the word subgraph for an induced subgraph. We write H⊆G. Let G be a graph and H⊆G. G-H will denote a subgraph of G obtained from G by deleting all points belonging to H. For two graphs G=(X,U) and H=(Y,W) with disjoint point sets X and Y, their join G⊕H is the graph that contains G, H and edges joining every point in G with every point in H. The subgraph O(v)⊆G containing all points adjacent to v (without v) is called the rim or the neighborhood of point v in G, the subgraph $U(v)=v\oplus O(v)$ is called the ball of v. Graphs can be transformed from one into another in a variety of ways. Contractible transformations of graphs seem to play the same role in this approach as a homotopy in algebraic topology. It was shown that contractible transformations retain the Euler characteristic and the homology groups of a graph [10-12].

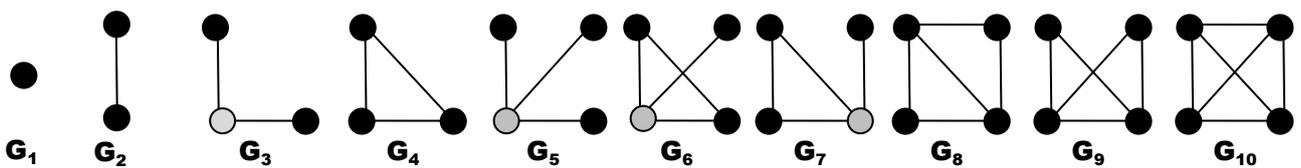

**Figure 4. Contractible graphs. Black points are simple.**

**Definition 3.1.**
   (a) The trivial graph $K_1$ (with only one point) is contractible.
   (b) Let G be a contractible graph and H be a contractible subgraph of G. The graph $G\cup\{v\}$ obtained by gluing a point v to G in such a way that the rim $O(v)=H$ is a contractible graph.
   (c) We say that x is a simple point of a graph G if the rim $O(x)$ of x a contractible graph.

Figure 4 shows contractible graphs with the number of points k<5. Black point of graphs are simple. Consider properties of contractible graphs.

**Proposition 3.1** ([10-12]).
Let G be a contractible graph.
- G can be converted to a point $v\in G$ by sequential deleting simple points.
- If $H\subset G$ is a contractible subgraph of G then G can be converted to H by sequential deleting simple points ( i.e., G contains a simple point belonging to G-H).
- Let G and H be graphs. If G is a contractible graph then $H\oplus G$ is a contractible graph.

In fig. 4, any graph $G_i$ can be converted to a one-point graph by sequential deleting simple points.

**Definition 3.2.**
- Deletions and attachments of simple points and edges are called contractible transformations.
- Graphs G and H are called homotopy equivalent if one of them can be converted to the other one by a sequence of contractible transformations.

Graph G(M) depicted in fig. 7 can be transformed to either graph $G(M_1)$ or graph $G(M_2)$ by sequential deleting simple points. Therefore, graphs G(M), $G(M_1)$ and $G(M_2)$ are homotopy equivalent. It follows from the previous results that if graphs G and H are homotopy equivalent, and G is contractible, then so does H.

**3. Simple n-cubes, contractible cubical spaces and homotopy equivalent cubical spaces**

In connection with computer experiments, let us consider homotopy properties of cubical spaces.

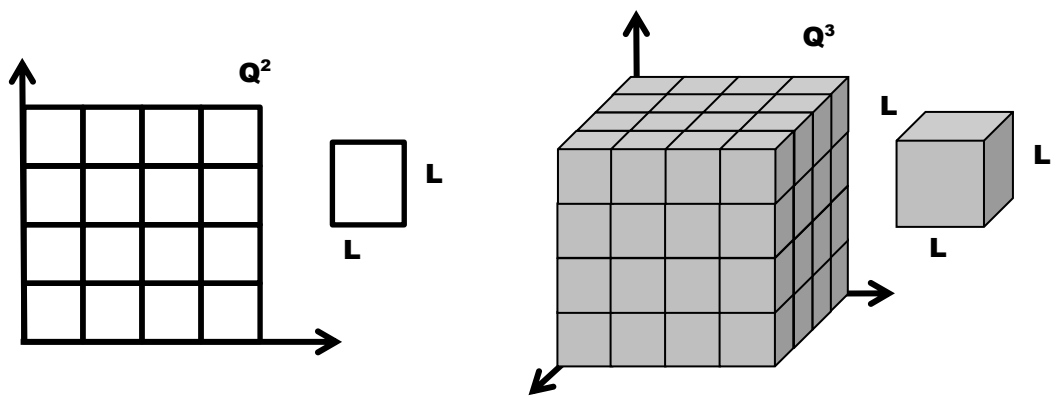

**Figure 5. Euclidian cubical n-spaces $Q^2$ and $Q^3$.**

In this paper, the Euclidian cubical n-space $Q^n$ is constructed as the set of n-cubes in Euclidean space $E^n$ of side length L and corner coordinates in the set $X=\{Lx_1,...Lx_n: x_i\in Z\}$. By construction, $Q^n$ partitions $E^n$.

Note, we do not consider cubical or abstract n-comlexes which are traditionally used in order to provide a topological background for digital images. A cubical space is a set of elements having the same size and dimension n, i.e., of n-cubes. Figure 5 shows Euclidian cubical n-spaces $Q^2$ and $Q^3$.

We say that M is a cubical space if M is a subset of $Q^n$. Evidently, if n-cubes $u_i$ and $u_k$ belong to M and $u_i \cap u_k \neq \emptyset$ then $u_i \cap u_k$ is a p-dimensional face of $u_i$ and $u_k$, i.e., a p-cube, $0 \leq p \leq n-1$. The union $I(M) = \cup \{u_i \mid u_i \in M\}$ of all n-cubes of M is called the image of M in $E^n$. Cubical spaces are shown in figures 6-8.

The subspace $O(u_i)$ of a cubical space $M = \{u_1, u_2, \ldots\}$ containing all n-cubes intersecting $u_i$ is called the rim or the neighborhood of $u_i$ in M, the cubical subspace $U(u_i) = u_i \cup O(u_i)$ is called the ball of $u_i$.

Let $M = \{u_1, u_2, \ldots\}$ be a cubical space. Then the intersection graph G(M) of M with points $\{x_1, x_2, \ldots\}$ is called the digital model of M if points $x_k$ and $x_i$ are adjacent whenever $u_k \cap u_i \neq \emptyset$. In other words, f: G(M)→M such that $f(x_i) = u_i$ is an isomorphism.

Cubical spaces and their digital models are shown in figures 6-8. Contractible cubical spaces and simple n-cubes are defined below by using a recursive definition.

**Definition 4.1.**
- A cubical space $M = \{u_1, u_2, \ldots\}$ is called contractible if it can be converted to an n-cube by sequential deleting simple n-cubes.
- An n-cube $u_i$ of a cubical space $M = \{u_1, u_2, \ldots\}$ is said to be simple if its rim $O(u_i)$ is a contractible cubical space.

Figure 6 shows contractible cubical spaces and simple cubes. Spaces $M_1$, $M_2$ and $M_3$ contain simple cube {a} because the rim O(a)={b} is a contractible one-cube space N. Since $M_i - \{a\}$, i=1,2,3, is a contractible space N={b} then $M_i$ is a contractible space. For the same reason, {a} is a simple cube in $M_4$ and $M_5$, and $M_4 - \{a\}$ and $M_5 - \{a\}$ are contractible cubical spaces. Therefore, $M_4$ and $M_5$ are contractible cubical spaces.

Cubical space M depicted in fig. 7 contains simple cubes a, b, c, h and e. The following corollary is a direct consequence of definition 4.1.

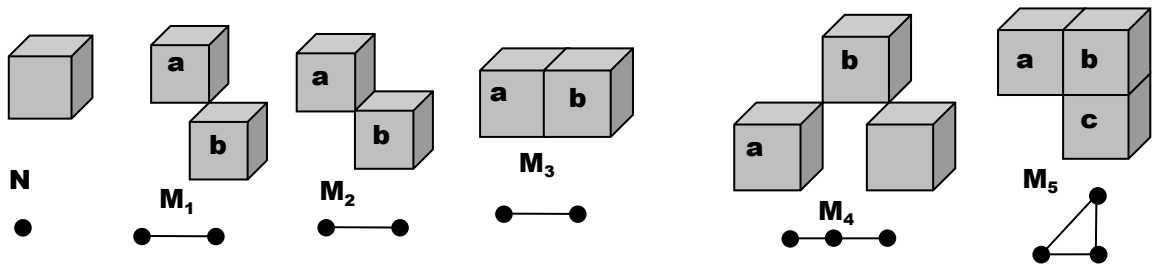

Figure 6. Contractible cubical spaces containing simple cube {a}.

**Corollary 4.1.**
Let M be a contractible cubical space. If u is a simple n-cube of M then the cubical space M-{u} is contractible.

**Proposition 4.1.**
If M is a contractible cubical space then the image I(M) has the homotopy type of a point in the framework of algebraic topology.

**Proof.**
The proof is by induction on the number |M| of n-cubes of M. For |M|=1,2, 3,… the proposition is plainly true. For example, images of all cubical spaces depicted in fig. 6 can be continuously shrunk to a point. Assume that the proposition is valid whenever |M|<k. Let |M|=k.

(a) Since M is a contractible cubical space, it contains a simple n-cube x, i.e., the rim O(x) is a contractible cubical space, the ball $U(x) = O(x) \cup x$ is a contractible cubical space and M-x is a contractible cubical space according to definition 4.1. With no loss of generality, suppose that |O(x)|<k-1. Then O(x) is a deformation retract of both $O(x) \cup x$ and M-{x}, i.e., $I(O(x) \cup x)$ and I(M-{x}) can be deformed continuously to I(O(x)).

This means that I(M) can be deformed continuously to I(O(x)). Since I(O(x)) has the homotopy type of a point then I(M) has the homotopy type of a point. . □

**Definition 4.2.**
- Cubical spaces M and N are called homotopy equivalent if one of them can be converted to the other one by sequential deleting and attaching simple n-cubes.
- A cubical space is compressed if it contains no simple n-cube.

A compressed contractible cubical space M contains only one n-cube. In fig. 7, cubical spaces M, $M_1$ and $M_2$ are homotopy equivalent. M can be converted to $M_1$ by sequential deleting simple n-cubes {c,b,a,h}, and to $M_2$ by sequential deleting simple n-cubes {c,b,e}. Cubical spaces $M_1$ and $M_2$ are compressed since they do not contain simple points.

**Proposition 4.2.**
Let cubical spaces M and N are homotopy equivalent. Then images I(M) and I(N) are homotopy equivalent topological spaces.
**Proof.**
It is necessary to show that deleting (attaching) a simple n-cube does not change the homotopy type of the image of a cubical space. Let M={u,v,…) be a cubical space and v be a simple n-cube in M. This means that the rim O(v) of v and the ball U(v)= O(v)∪v are a contractible cubical spaces, i.e., the image I(O(v)∪v) can be deformed continuously to I(O(v)) according to proposition 4.1. Hence, the image I(M) can be deformed continuously to the image I(M-v). Therefore, I(M) and I(M-v) are homotopy equivalent. For the same reason, if x is a simple n-cube in M∪x then I(M) is homotopy equivalent to I(M)∪x). After repeating this process for each n-cube in the sequence of transformations, we obtain that images I(M) and I(N) are homotopy equivalent spaces in the framework of algebraic topology. □

The images I(M), I($M_1$) and I($M_2$) in fig. 7 are homotopy equivalent to a circle. The following corollary summarizes previous results.

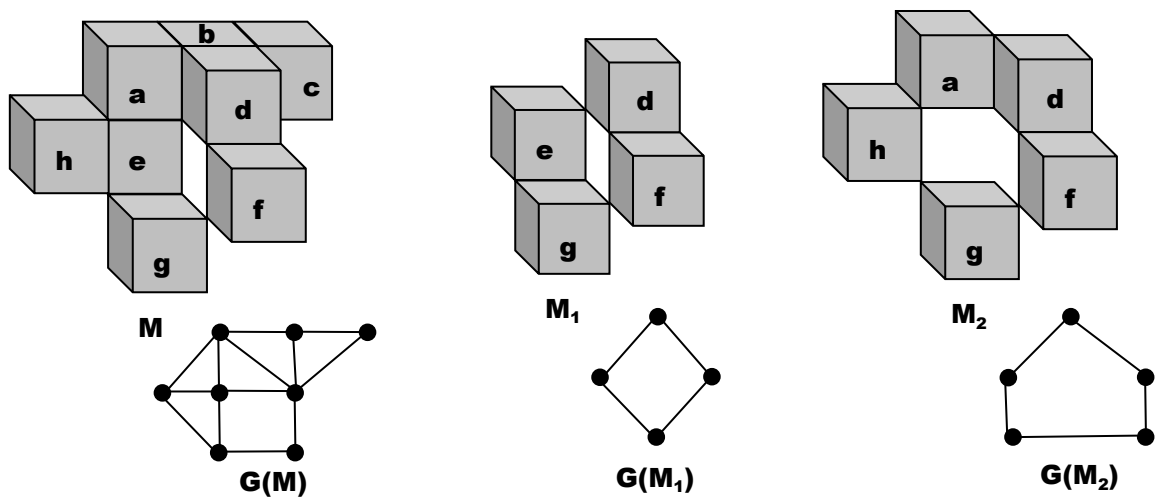

**Figure 7.  Cubical spaces M, $M_1$ and $M_2$ are not contractible. $M_1$ and $M_2$ are compressed cubical spaces. . G($M_1$) and G($M_2$)  are compressed graphs.**

**Corollary 4.2.**
Let M be a cubical space and N be a compressed cubical space obtained from M by sequential deleting simple n-cubes. Then images I(M) and I(N) are homotopy equivalent spaces.

In general, compressed forms of a given cubical space are not isomorphic. Compressed cubical spaces $M_1$ and $M_2$ in fig. 7 are obtained from M and homotopy equivalent but not isomorphic.

The following proposition shows a close link between homotopy properties of cubical spaces and their intersection graphs.

**Proposition 4.3.**
Let $M=\{u_1,u_2,...\}$ be a cubical space and $G(M)=\{x_1,x_2,...\}$ be the intersection graph of M. If M is a contractible cubical space then G(M) is a contractible graph.

**Proof.**
The proof is by induction on the number |M| of points of M. For |M|=1,2, 3,… the proposition is plainly true. For example, for all contractible cubical spaces depicted in fig. 6, their intersection graphs are contractible.

Assume that the proposition is valid whenever |M|<k. Let |M|=k. Since M is a contractible cubical space, it contains a simple n-cube u, i.e., the rim O(u) is a contractible cubical space, the ball U(u)= O(u)∪u is a contractible cubical space and M-u is a contractible cubical space according to definition 4.1. With no loss of generality, suppose that |O(u)|<k-1.
Let x be the point of G(M) associated to u, O(x) be the rim of x, U(x)=O(x)∪x be the all of x, and G(M)-x be the graph obtained by deleting x from G(M). The intersection graphs G(O(u))=O(x), G(U(u))=U(x) and G(M-u)=G(M)-x are all contractible graphs by the induction hypotheses. Therefore, G(M) is a contractible gaph accorging to definition 3.1. □

Corollary 4.3 directly follows from propositions 3.1 and 4.3.

**Corollary 4.3.**
Let M be a contractible cubical space.
- If |M|>1 then M contains at least two simple n-cubes.
- If N⊂M is a contractible cubical subspace of M then M can be converted to N by sequential deleting simple n-cubes.

Let cubical spaces M and N are homotopy equivalent. Then the intersection graphs G(M) and G(N) are homotopy equivalent graphs/

Fig. 7 shows cubical spaces M, $M_1$, and $M_2$ and the intersection graphs G(M), G($M_1$) and G($M_2$). M can be converted to $M_1$ by sequential deleting simple n-cubes {c,b,a,h}, and to $M_2$ by sequential deleting simple n-cubes {c,b,e}. It is easy to see that G(M) can be converted to G($M_1$) by sequential deleting simple points corresponding to n-cubes {c,b,a,h}, and can be converted to G($M_2$) by sequential deleting simple points corresponding to n-cubes {c,b,e}. Cubical space M in fig. 8 is a cubical model of a sphere (as in fig. 3). $M_1$, $M_2$ and $M_3$ are cubical spaces obtained from M by sequential deleting simple 3-cubes. $M_3$ is a compressed cubical space. G($M_1$), G($M_2$) and G($M_3$) are intersection graphs of $M_1$, $M_2$ and $M_3$ sequentially obtained from G(M) by sequential deleting simple 3-cubes. G($M_3$) is a minimal digital 2-sphere [12]. Clearly, graphs G(M), G($M_1$), G($M_2$) and G($M_3$) are homotopy equivalent.

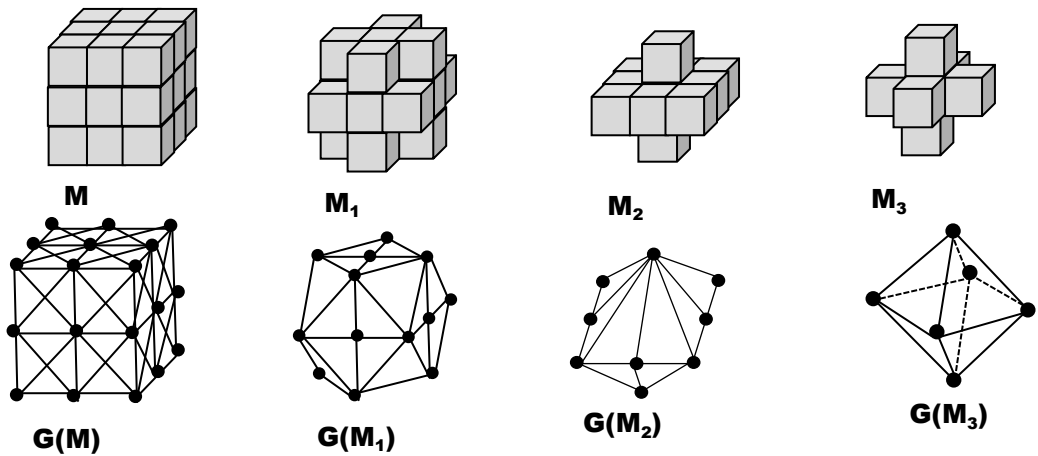

**Figure 8. M, $M_1$, $M_2$ and $M_3$ are homotopy equivalent cubical spaces. Digital models G(M), G($M_1$), G($M_2$) and G($M_3$) are homotopy equivalent graphs.**

Notice that cubical spaces may serve as useful tools for building detailed digital models of 2-, 3- and k-dimensional spaces and surfaces in Euclidean space $E^n$ including such spaces as the projective plane, the Bing's house and the dunce hat.

Now we can describe a method for constructing cubical and digital models preserving basic topological properties of continuous objects.

- In Euclidean space $E^n$, build the cubical space Q={$u_1,u_2,…$} as the set n-cubes in $E^n$ of side length L and corner coordinates in the set X={ $Lx_1,…Lx_n$: $x_i \in Z$}, provided that L is small enough. This means that starting from a rough initial side length L, L is iteratively selected guided by the quality of the approximation.
- For a continuous object W, construct the cubical model M={$u_1,u_2,…$} with the following properties: W⊆ ∪{$u_i$ |$u_i \in M$}, and if $u_i \in M$ then $u_i \cap W \neq \emptyset$.
- Sequentially delete simple n-cubes from M. We obtain a compressed cubical model N of W.
- Construct the digital models G(M) and G(N) of M and N that are their intersection graphs.

We can work with either cubical spaces or their intersection graphs. It is clear, that realizing this method is a computational problem, which is understood to be a task that is in principle capable to being solved by a computer.

**Conclusion**

- N-dimensional objects and their cubical models are homotopy equivalent spaces in the framework of algebraic topology.
- A cubical space can be converted to a compressed form by sequential deleting simple cubes.
- Deleting and attaching a simple cube preserves the homotopy type of a cubical space.
- If cubical spaces are homotopy equivalent then their intersection graphs are homotopy equivalent.